\documentclass[aps,preprint,superscriptaddress,11pt]{revtex4}

\usepackage{amsmath}
\usepackage[dvips]{graphics}
\usepackage[dvips]{graphicx}
\usepackage{psfrag}
\usepackage[mathcal]{eucal}

\begin{document}


%{\large 

\preprint{BNL-HET-05-14}

\title{Generation of Small Neutrino Majorana Masses 
%with a Scalar $SU(2)$-triplet 
in a Randall-Sundrum Model}


\author{Mu-Chun Chen}
\email[]{chen@quark.phy.bnl.gov}
\affiliation{High Energy Theory Group, Department of Physics, 
Brookhaven National Laboratory, Upton, 
NY 11973-5000}


%\date{\today}

\begin{abstract}
We propose a model, in the framework of $5D$ with 
warped geometry, in which small neutrino Majorana 
masses are generated by tree level coupling  
of lepton doublets to a $SU(2)_{L}$-triplet 
scalar field, which is coupled to a bulk SM-singlet. 
The neutrino mass scale is determined by the bulk mass term $(\alpha_{S})$ 
of the singlet as $ve^{-2(\alpha_{S}-1)\pi kR}$. 
This suppression is due to a small overlap between the 
profile of the singlet zero mode and the triplet, 
which is confined to the TeV brane. 
The generic form for the neutrino mass matrix due to the 
overlap between the fermions is not compatible with the 
LMA solution. This is overcome by imposing a $Z_{4}$ 
symmetry, which is softly broken by couplings of the 
triplet Higgs to the lepton doublets. This model successfully 
reproduces the observed masses and mixing 
angles in charged lepton sector as well as in the neutrino 
sector, in addition to having a prediction of 
$\left| U_{e3} \right| \sim \mathcal{O}(0.01)$. 
The mass of the triplet is of the order of a TeV, 
and could be produced at upcoming collider experiments.
The doubly charged member of the triplet can decay into two 
same sign charged leptons yielding the whole triplet coupling matrix 
which, in turn, gives the mixing matrix in the neutrino sector. 
%The singlet zero mode, which has a TeV brane induced 
%mass of the order of an MeV, set by the neutrino mass scale, 
%is weakly interacting, 
%and thus it is a natural candidate for the dark matter. 
\end{abstract}

%\keywords{}

\maketitle


\section{Introduction}\label{introduction}

Many new ideas aiming to solve the gauge hierarchy problem have 
emerged in recent  
years. Randall and Sundrum (RS)~\cite{Randall:1999ee} have 
proposed a solution based on non-factorizable geometry 
in a slice of $AdS_{5}$ space with warped background metric. 
The warped metric is a solution to the Einstein equation imposing the $4D$ 
Poincare invariant,
\begin{equation}\label{RSmetric}
ds^{2} = e^{-2\sigma(y)} \eta_{\mu\nu} dx^{\mu} dx^{\nu} + dy^{2} \; ,
\end{equation}
with a scale factor 
\begin{equation}
\sigma(y) = k \left| y \right|  \; ,
\end{equation}
where $y$ is the 5-th coordinate, $\eta_{\mu\nu}$ is the metric in $4D$ flat 
space given by
$\mbox{diag} \left( -1,1,1,1 \right)$, and $k$ is a parameter related
 to the Ricci scalar and bulk 
vacuum energy. It is naturally of the order of the $5D$ Planck scale,
 $M_{pl}$.  
Throughout this paper we adopt the convention that 
Greek letters $\left( \mu,\nu, ... \right)$ refer to $4D$ space-time
 indices while Latin letters 
$(A,B, ...)$ refer to $5D$ indices. 
The space along the $5$-th dimension is $S^{1}/Z_{2}$. 
Two $3$-branes are located at the orbifold fixed points: the Planck-brane 
at $y=0$ and the TeV-brane at $y= \pi R$. 

Consider the SM Higgs field confined to the TeV brane,
the five-dimensional action for the scalar sector reads,
\begin{equation}\label{higgspot}
\int d^{4}x \int dy \; \sqrt{-g} \; \delta \biggl( y-\pi R \biggr) 
\biggl[ g^{\mu\nu} D_{\mu} H(x)^{\dagger} D_{\nu} H(x)
-\lambda \biggl(  \left| H(x) \right|^{2} - v_{5}^{2} \biggr)^{2}
 \biggr]  \; , 
\end{equation}
where $\sqrt{-g}$ is defined as $\sqrt{-g} \equiv 
\sqrt{-\mbox{det} g_{AB}}$, 
which is equal to $e^{-4\sigma(y)}$ for the RS metric; 
$v_{5}$ is the vacuum expectation value 
(VEV) of the Higgs, 
thus the symmetry breaking scale in $5D$. 
After integrating out the 5-th coordinate, the kinetic term for $H(x)$ has 
a coefficient $e^{-2\pi kR}$. In order to get the canonically normalized 
kinetic term for $H(x)$, we must perform the following rescaling 
$H(x) \rightarrow e^{\pi k R} \widetilde{H}(x)$, and identify 
$\widetilde{H}(x)$ as the 
physical Higgs field. The resulting effective $4D$ action with canonically 
normalized kinetic term then reads
\begin{equation}\label{effhiggspot}
\int d^{4} x \;
\eta^{\mu\nu}D_{\mu}\widetilde{H}(x)^{\dagger}D_{\nu}\widetilde{H}(x) -
\lambda \left(
|\widetilde{H}(x)|^{2} - e^{-2\pi kR} v_{5}^{2}
\right)^{2} \; .
\end{equation}
The symmetry breaking scale in the $4D$ effective theory, $v$, 
is therefore related to the $5D$ symmetry breaking scale by
\begin{equation}
v = e^{-\pi kR} v_{5} \; .
\end{equation}
Note that in the $5D$ action of Eq.~(\ref{higgspot}), 
$v_{5}$ has mass dimension $1$, 
thus its natural value is of the order of $M_{pl}$. 
The $4D$ Planck scale, 
$\overline{M}_{pl}$, is obtained after integrating out 
the fifth coordinate in the gravitational action, and is given by
\begin{equation}\label{eq:reducedM}
\overline{M}_{pl}^{2} = \frac{M_{pl}^{3}}{k} (1-e^{-2\pi kR}) \; .
\end{equation}
Thus the electroweak scale, $v$, 
and the $4D$ Planck scale are related by
\begin{equation}
\frac{v}{\overline{M}_{pl}} = e^{-\pi kR} \sqrt{\frac{k}{M_{pl}}}
\sqrt{\frac{1}{(1-e^{-2\pi kR})}}
\; \stackrel{kR \gg 1}{\simeq} \; e^{-\pi kR} \sqrt{\frac{k}{M_{pl}}} \; .
\end{equation}
Assuming $k=M_{pl}$, with the choice of $kR \approx 11$, we see that 
the TeV electroweak symmetry breaking scale, $v$, can be derived from 
the warped factor $e^{-\pi kR}$. Thus the gauge hierarchy 
between $4D$ electroweak symmetry breaking scale and the $4D$ Planck scale 
is resolved. In what follows, the SM Higgs doublet will be the only field 
confined to the TeV-brane; all other fields are allowed to 
propagate in the $5$-th dimension, unless otherwise 
stated~\cite{Gherghetta:2000qt,Huber:2000ie}. 

The conventional way to generate neutrino masses in $4D$ is the 
see-saw mechanism. In this mechanism, the smallness of 
the neutrino masses is related to the large 
mass scale (typically of the order of the grand unification scale)
 of the right-handed neutrinos. A natural framework 
to accommodate the see-saw mechanism is $SO(10)$. 
For reviews, see for example, Ref.~\cite{Chen:2000fp,Chen:2003zv}.
In the Randall-Sundrum scenario, the warped geometry provides 
new ways to generate the fermion mass hierarchy.  
Models with RS geometry have been constructed to naturally 
accommodate small Dirac 
neutrino masses~\cite{Grossman:1999ra,Huber:2000ie,Huber:2001ug}. 
To see how it works, let us first consider the Yukawa interaction in $5D$,
\begin{equation}
\frac{Y_{r}}{M_{pl}} \int d^{4}x \int dy \sqrt{-g} \;
 \overline{\Psi}_{R}(x,y) 
\Psi_{L}(x,y) H(x) \delta(y-\pi R) \; ,
\end{equation}
where $r = e, \nu, ... \mbox{etc.}$ 
The $4D$ effective Yukawa coupling is obtained after integrating out the 
5-th coordinate~\cite{Gherghetta:2000qt},
\begin{equation}
\widetilde{Y}_{r} = \frac{Y_{r}}{M_{pl}} \frac{1}{2\pi R}
\biggl[
\frac{(1-2c_{R,r})\pi kR}{e^{(1-2c_{R,r})\pi kR} -1 }
\biggr]^{1/2}
\biggl[
\frac{(1-2c_{L,r})\pi kR}{e^{(1-2c_{L,r})\pi kR} -1 }
\biggr]^{1/2}
e^{(1-c_{L,r}-c_{R,r}) \pi kR} \; ,
\end{equation}
where $c_{L,r}$ and $c_{R,r}$ parameterize the bulk mass terms
 of the fermions. 
Detailed derivations of the above results can be found in the 
following sections. 
(Throughout the paper, the $5D$ coupling constants are un-tilded, while 
the $4D$ effective coupling constants are tilded.) 
It is clear that to generate small Dirac masses for neutrinos requires 
\begin{equation}\label{eq:ratio}
\frac{m_{\nu}}{m_{e}} = \frac{\widetilde{Y}_{\nu}}{\widetilde{Y}_{e}} 
\frac{ \bigl< H(x) \bigr> }{ \bigl< H(x) \bigr> } \sim
\biggl[
\frac{(1-2c_{R,\nu})\pi kR}{e^{(1-2c_{R,\nu})\pi kR} -1 }
\biggr]^{1/2}
\biggl[
\frac{e^{(1-2c_{R,e})\pi kR} -1 }{(1-2c_{R,e})\pi kR}
\biggr]^{1/2}
\frac{e^{-c_{R,\nu}\pi k R}}{e^{-c_{R,e}\pi k R}} \ll 1  \; .
\end{equation}
This inequality can be satisfied if $c_{R,e} < c_{R,\nu}$. 
Therefore, if the right-handed neutrino is localized 
closer to the Planck brane than the right-handed charged lepton is, 
the smallness of the neutrino masses compared to the charged lepton 
masses can be explained. 

If the neutrinoless double beta decay is observed, neutrino masses 
must be of the 
Majorana type. Nevertheless, it is difficult to generate small 
Majorana masses 
for the neutrinos through the conventional see-saw mechanism 
with warped geometry.  
To see how this comes about, consider the following operator
\begin{eqnarray}
& \int d^{4}x \int dy \sqrt{-g} \frac{\lambda_{ij}}{M_{\mbox{ \tiny pl}}^{2}} 
H(x)^{2} \Psi_{L,i}^{T}(x,y) \mathcal{C} \Psi_{L,j}(x,y) \delta(y-\pi R)
\nonumber\\
& \equiv \int d^{4}x \widetilde{M}_{ij} \Psi_{L,i}^{(0) \, T}(x)
\mathcal{C} \Psi_{L,j}^{(0)}(x) 
+ \cdots \; ,
\end{eqnarray}
where $\mathcal{C}$ is the charge conjugation operator, 
$\cdots$ denotes terms involving higher level KK modes
 (throughout this paper), 
and the four-dimensional effective Majorana mass matrix is given by
\begin{eqnarray}
\widetilde{M}_{ij} & = & 
\frac{1}{2\pi R} \frac{\lambda_{ij}}{M_{pl}^{2}}
\left[
\frac{(1-2c_{L,i})\pi kR}{e^{(1-2c_{L,i})\pi kR} -1 }
\right]^{1/2}
\left[
\frac{(1-2c_{L,j})\pi kR}{e^{(1-2c_{L,j})\pi kR} -1 }
\right]^{1/2}
\nonumber\\
& & \qquad \cdot
\int_{-\pi R}^{\pi R} dy 
e^{-4\sigma(y)} e^{(2-c_{L,i})\sigma} 
e^{(2-c_{L,j})\sigma} 
e^{2\sigma(y)} \delta(y-\pi R) \cdot v^{2}
\nonumber\\
& = &
\frac{1}{2 \pi R} \frac{\lambda_{ij}}{M_{pl}^{2}}
\left[
\frac{(1-2c_{L,i})\pi kR}{e^{(1-2c_{L,i})\pi kR} -1 }
\right]^{1/2}
\left[
\frac{(1-2c_{L,j})\pi kR}{e^{(1-2c_{L,j})\pi kR} -1 }
\right]^{1/2}
e^{(2-c_{L,i}-c_{L,j})\pi kR} \cdot v^{2} \; .
\end{eqnarray}
The only way to have $\widetilde{M}_{ij}$ suppressed 
and generate the correct mass scale for 
the neutrinos is when $(c_{L,i}+c_{L,j})$ is close to $2$.  
However, this condition leads to unrealistically small 
charged fermion masses~\cite{Huber:2002gp}. 
Therefore, $\lambda_{ij}$ 
must be extremely small in order to give small neutrino Majorana masses. 
Even with various mechanisms, including lowering $k/M_{pl}$ to $O(0.01)$ 
and having very strong thus non-perturbative $5D$ Yukawa
 coupling ($Y/g \sim O(10)$ where 
$g$ is the $5D$ gauge coupling constant) for the charged fermions,  
as mentioned in Ref.~\cite{Huber:2002gp}, to bring down the Majorana masses, 
a tiny value as small as $10^{-4}$ in natural units 
for the coupling constant is still needed. It is the aim of our paper 
to provide a new mechanism that gives rise to small Majorana masses 
in a more natural way,  
using parameters all of order unity in natural units. We will show that 
this is possible if the Majorana masses are generated by coupling 
the lepton doublets to a $SU(2)$-triplet Higgs.

The next section contains preliminaries concerning the equations of motion 
of bulk fields and their zero-mode solutions. Our model with a 
$SU(2)$-triplet is described in Section III. 
Numerical results are in Section IV and conclusions 
in Section V.

\section{Localization of bulk fields}\label{localization}

The equations of motion for various bulk fields are given 
in the following compact form~\cite{Grossman:1999ra,Gherghetta:2000qt}
\begin{equation}\label{geneom}
(e^{2\sigma}\eta^{\mu\nu}\partial_{\mu}\partial_{\nu}
+ e^{s\sigma}\partial_{5}(e^{-s\sigma}\partial_{5}) - M_{\Phi}^{2})
\Phi(x^{\mu},y) = 0 \; ,
\end{equation}
where for $\Phi=(\phi,e^{-2\sigma}\Psi_{L,R})$ we have 
$M_{\Phi}^{2}=(ak^{2}+b\sigma^{''}(y),C(C \pm 1)k^{2} 
\pm C\sigma^{''}(y))$ and $s=(4,1)$, where $a$ and $C$ are bulk mass terms 
of the scalar and fermionic fields, and $b$ is a boundary mass
 term for the scalar 
field. Due to the presence of these bulk mass terms, components of 
the bulk field in 4D 
can develop profiles that depend on the fifth coordinate $y$.   
Decompose the field $\Phi(x^{\mu},y)$ into an infinite sum of
Kaluza-Kline (KK) modes
\begin{equation}\label{KKdecompose}
\Phi(x^{\mu},y)=\frac{1}{\sqrt{2\pi R}} \sum_{n} 
\Phi_{(n)} (x^{\mu}) f_{n}(y) \; .
\end{equation}
The profile of the n-th mode, $f_{n}(y)$, satisfies
\begin{equation}\label{eomfn}
(-e^{s\sigma}\partial_{5}(e^{-s\sigma} \partial_{5})+\hat{M}_{\Phi}^{2})
f_{n}(y) = e^{2\sigma} m_{n}^{2} f_{n}(y)  \; ,
\end{equation}
where $\hat{M}_{\Phi}^{2} = (ak^{2}, C(C \pm 1)k^{2})$. 
$m_{n}$ is the mass of the $n$-th KK
mode. Solutions for the zero modes of various bulk fields have been found 
previously~\cite{Gherghetta:2000qt,Grossman:1999ra}. 
We summarize the results in the following:  

{\it Spin-0 fields:} 
If boundary mass terms with $b=\bigl(2-\alpha\bigr)$ are present, 
a zero mode solution ($m_{0}=0$) for the scalar field can exist. 
The relation between these two parameters $b$ and $\alpha$ can be 
justified in the SUSY limit~\cite{Gherghetta:2000qt}. 
The zero mode solution is given by 
\begin{equation}\label{scalar}
f_{0}(y) = \frac{1}{N_{0}} e^{(2-\alpha)\sigma} \; ,
\end{equation}
where $\alpha=\sqrt{4+a}$ and $a$ is the bulk scalar mass.
The normalization constant $1/N_{0}$ is given by
\begin{equation}
\frac{1}{N_{0}^{2}} = \frac{(2-2\alpha)
 \pi k R}{e^{(2-2\alpha)\pi k R} -1} \; .
\end{equation}
Thus the bulk scalar field can be decomposed into
\begin{equation}
\phi(x^{\mu},y) = \frac{1}{\sqrt{2\pi R}} \frac{1}{N_{0}}
 e^{(2-\alpha)\sigma} 
\phi_{(0)}(x^{\mu}) + \cdots \; .
\end{equation}
For $\alpha = 1$, the normalization factor $1/N_{0}=1$ 
and the zero mode profile  
is $e^{-\sigma(y)}f_{0}=1$, 
where the factor $e^{-\sigma(y)}$ accounts for the 
non-trivial measure due to the warped geometry. 
In other words, the zero mode is de-localized, 
and we recover the conformal limit. 
For $\alpha > 1$ ($\alpha < 1$), 
the normalization factor is of order $1$ 
($e^{-(1-\alpha)\pi kR}$), and the zero mode is localized toward 
the Planck brane (TeV brane).

{\it Spin-1/2 fields:}
The solution in this case is
\begin{equation}\label{LH}
f_{0}(y) = \frac{1}{N_{0}} e^{-c\sigma} \; ,
\end{equation}
where $c=C$ for left-handed fermions and $c=-C$ for right-handed fermions. 
The normalization constant $1/N_{0}$ is given by
\begin{equation}
\frac{1}{N_{0}^{2}} = \frac{(1-2c) \pi k R}{e^{(1-2c)\pi k R} -1} \; .
\end{equation}
Thus the bulk fermion can be decomposed into
\begin{equation}
\Psi_{L}(x^{\mu},y) = e^{2\sigma} \Phi(x^{\mu},y) 
= \frac{1}{\sqrt{2\pi R}} \frac{1}{N_{0}} e^{(2-c)\sigma} 
\Phi_{L(0)}(x^{\mu}) + \cdots \; .
\end{equation}
For $c = 1/2$, the normalization factor $1/N_{0}=1$ and the
 profile of the zero mode 
$e^{-\frac{3}{2}\sigma(y)}( e^{2\sigma(y)} f_{0}(y)) = 1$ 
taking into account the measure $e^{-\frac{3}{2}\sigma(y)}$
 due to warped geometry.
For $c > 1/2$ ($c < 1/2$), the normalization factor is of order $1$ 
($e^{-(1/2-c)\pi kR}$), and the zero mode is localized toward 
the Planck brane (TeV brane).

\section{Small Majorana masses}\label{Majorana}

Now we introduce a $SU(2)$-triplet Higgs, $T$, which carries
 hypercharge $Y=2$ and 
lepton number $L=-2$, to generate Majorana masses for the
 left-handed neutrinos. 
$T$ can be written in terms of three component fields
\begin{equation}
T=\left(\begin{array}{cc}
\xi^{+}/\sqrt{2} & \xi^{++}\\
\xi^{0} & -\xi^{+}/\sqrt{2}
\end{array}
\right) \; .
\end{equation}
Its coupling to the lepton doublets is given by
\begin{equation}
\lambda_{ij} (L^{T}_{i} \mathcal{C}^{-1} i\tau_{2} T L_{j})
= \lambda_{ij} \biggl[
-\xi^{0} \nu_{L,i}^{T} \mathcal{C}^{-1} \nu_{L,j}
+ \frac{\xi^{+}}{\sqrt{2}} 
\biggl(\nu_{L,i}^{T} \mathcal{C}^{-1} l_{L,j} 
+ l_{L,i}^{T} \mathcal{C}^{-1} \nu_{L,j} \biggr)
+ \xi^{++} l_{L,i}^{T} \mathcal{C}^{-1} l_{L,j}\biggr] \; .
\end{equation}
If $T$ acquires VEV along the $\xi^{0}$ direction, the first
 term in the above 
equation then generates Majorana masses for $\nu_{L}$, 
\begin{equation}\label{triplet}
\lambda_{ij} <\xi^{0}> \nu_{L,i} \nu_{L,j}  \; .
\end{equation}
To have small neutrino masses compared to the charged lepton masses 
thus requires either ({\it i})
$\lambda_{ij}$ is much smaller than the Yukawa coupling of the
 charged leptons, 
or ({\it ii}) $<\xi^{0}>$ is much smaller than the electroweak symmetry 
breaking scale, or a combination of both.
Nevertheless, the first possibility is constrained by the experimentally 
well-measured 
$\rho \equiv \frac{M_{W}^{2}}{M_{Z}^{2}\cos^{2}\theta_{W}}
 \simeq 1$ relation. 
In the SM, $\rho=1$ is predicted. In the presence of one $SU(2)$ doublet
and one $SU(2)$ complex triplet with $Y=2$, this relation is 
modified to be
\begin{equation}
\rho = \frac{v^{2} + 2 v_{\xi^{0}}^{2}}{ v^{2} + 4 v_{\xi^{0}}^{2}}
= 1- \frac{2v_{\xi^{0}}^{2}}{v^{2} + 4v_{\xi^{0}}^{2}}
\equiv 1 + \Delta \rho \; ,
\end{equation}
where $<\xi^{0}> \equiv v_{\xi^{0}} / \sqrt{2}$. 
Experimentally, the $2\sigma$ level limits are,
\begin{equation}
-1.7 \times 10^{-3} < \Delta \rho <  2.7 \times 10^{-3} \; .
\end{equation}
This translates into the following bound on $v_{\xi^{0}}$~\cite{Haber:1999zh},
\begin{equation}
v_{\xi^{0}} < v / (24 \sqrt{2}) \; .
\end{equation}
A more detailed study utilizing a 
consistent renormalization scheme in the presence of a $SU(2)_{L}$ has been 
discussed in Ref.~\cite{Chen:2003fm}. 
We have found that this bound can be satisfied with mild
 fine-tuning of various parameters 
about $1$ part in $100$. However, because our philosophy is to restrict 
most parameters to be strictly of the order unity, 
we utilize the second possibility in 
our model. The second possibility has been utilized to generate small 
neutrino masses in $4D$ (See for example, Ref.~\cite{Ma:1998dx})
 and in $5D$ with 
a large extra dimension~\cite{Ma:2000wp}. It has not been 
considered before in $5D$ Randall-Sundrum model. 

In what follows, we propose a model that generates small neutrino Majorana 
masses. Consider the following potential for a scalar field 
$\phi$, $\phi^{4} + a m_{\phi}^{2} \phi^{2} + e\phi$. In the 
case of a $SU(2)$-triplet, 
$e$ is induced by the coupling to the SM Higgs doublet, $e = e^{'} <H>^2$,  
where $e^{'}$ is the coupling constant. It has been shown that $e$ must be 
non-zero in order to avoid the Majoron problem~\cite{Ma:1998dx}. 
If the square mass term, $am_{\phi}^{2}$ is positive and large 
(compared to $|e/am_{\phi}^{2}|$), the VEV for the scalar field 
is determined by the quadratic and the linear terms of the potential. 
%(see Fig.~\ref{potentialplot}). 
%\begin{figure}[t]
%\psfrag{V}[][]{$V$}
%\psfrag{T}[][]{$\phi$}
%\includegraphics[scale=0.4,angle=270]{potential.ps}
%\caption{\label{potentialplot} The dashed line denotes a potential 
%$\phi^{4} + a\phi^{2}$, while the solid line denotes a potential 
%with an added linear term $\phi^{4} + a\phi^{2} + e \phi$.}
%\end{figure} 
For $am_{\phi}^{2} \gg e^{'}v$, a small VEV for $\phi$ is generated. 
%In the RS model, 
There are two possible ways to satisfy this inequality: 
({\it i}) $am_{\phi}^{2}$ is much higher than the weak scale 
while $e$ scales as $v^{3}$; 
({\it ii}) $am_{\phi}^{2}$ is of the weak scale, $v^{2}$, while $e$ is 
highly suppressed with respect to $v^{3}$; that is, the 
coupling $e^{\prime} \ll v$. 
Due to the warped geometry, the relevant mass scale in 4D is $v$
 rather than $M_{pl}$.  
Thus the scalar mass term $m_{\phi}$ has to be of the order
 of the weak scale. 
This leaves us with the second choice, $e^{\prime} \ll v$,
 which can be naturally 
achieved utilizing the warped geometry. 

As it turns out, having a bulk triplet scalar 
does not work, as the induced triplet VEV in 
this case is of the order of the weak scale.  
As we show below, by confining the triplet scalar to the TeV brane and 
by introducing a bulk  SM singlet scalar field, $S(x,y)$, 
naturally small Majorana masses for the neutrinos can indeed arise.

{\bf The Model:} As we have mentioned above, the triplet is confined to the 
TeV brane along with the SM Higgs doublet. All other fields
 propagate in the bulk. 
The bulk Lagrangian for the free singlet scalar field is given by,
\begin{eqnarray}\label{5dfree}
\int d^{4}x \int dy \sqrt{-g} \biggl[ g^{\mu\nu} \bigl(D_{\mu}
 S(x,y)\bigr)^{\dagger}
\bigl(D_{\nu}S(x,y)\bigr) - a_{S} k^{2} S(x,y)^{2} \nonumber\\
\qquad - b_{S} k\bigl( \delta(y-0) - \delta(y-\pi R) \bigr) S(x,y)^{2}
 \biggr] \; .
\end{eqnarray}
For non-zero boundary mass terms with $b=\bigl(2-\alpha_{S}\bigr)$,
 where $\alpha_{S}=\sqrt{4+a_{S}}$, 
the singlet field has a massless zero mode, whose
 profile along the 5th dimension is determined by 
the parameter $\alpha_{S}$. We then add the interactions
 of the singlet scalar with the SM Higgs and the triplet
 on the TeV brane, described by the scalar potential,
 $V_{vis}$. As the couplings of these interactions are
 highly suppressed compared to the boundary mass terms given
 in Eq.~(\ref{5dfree}),  their effects on the profile of
 the singlet are negligible. 
The most general renormalizable (from the 4D point of view)
 potential which respects 
the $SU(2)_{L}$ symmetry is the following. 
\begin{eqnarray}\label{potential:vis}
V_{vis} & = & \delta(y-\pi R) \biggl[
\lambda_{H} H^{4}(x) + \mu_{H} k^{2} H^{2}(x) 
+\lambda_{T} T(x)^{4} + \mu_{T} k^{2} T(x)^{2} 
+\lambda_{S}S(x)^{4}
\nonumber\\
& & \qquad + \eta T(x)^{2} H(x)^{2}  
+ \frac{\chi_{1}}{M_{pl}} T(x)^{2} S(x,y)^{2}
+ \frac{\chi_{2}}{M_{pl}} H(x)^{2} S(x,y)^{2}
\nonumber\\
&&\qquad
+ \gamma_{1} \sqrt{M_{pl}} T^{2}(x) S(x,y)
+   \gamma_{2} \sqrt{M_{pl}} H^{2}(x) S(x,y)
+\zeta M_{pl} H(x) T(x) H(x)
\nonumber\\
&&\qquad 
+ \frac{\xi}{\sqrt{M_{pl}}} H(x) T(x) H(x) S(x,y) \; \biggr] 
+ h.c. \; ,
\end{eqnarray}
where $\lambda_{S}$, $\lambda_{i}$, $\mu_{i} \; (i=T,H)$, $\eta$, 
$\zeta$, $\chi_{i}$, $\gamma_{i}$ and $\xi$   
are dimensionless $\mathcal{O}(1)$ parameters.
Here we have used the following short-hand notations:
 $H^{4} = (H^{\dagger}H)^{2}$, 
$H^{2} = H^{\dagger}H$, $T^{4} = (Tr(T^{\dagger}T))^{2}$, $T^{2}
 = Tr(T^{\dagger}T)$,
$HTH = H^{\dagger} T H^{c}$ where $H^{c} = i \tau_{2} H^{\ast}$.
 For the quartic term, 
$T^{2}H^{2}$, there are two possible $SU(2)$ contractions: 
$H^{\dagger}H Tr(T^{\dagger}T)$ and 
$i \epsilon_{ijk} H^{\dagger} \sigma^{i} H T^{\dagger j} T^{k}$.
 Because these two terms 
are not important for our discussion, we do not distinguish 
them in the potential. 
Note that the couplings $HTH$ and $HTHS$ explicitly breaks 
the lepton number. After integrating out the fifth coordinate, 
we obtain the following effective 
coupling constants,
\begin{eqnarray}
\widetilde{\lambda}_{i} =
\lambda_{i} \int_{-\pi R}^{\pi R} dy \;
e^{-4\sigma} e^{4\sigma} \delta(y-\pi R)
= \lambda_{i} \qquad 
\nonumber\\
(\mbox{similarly for the quartic coupling $\eta$})
\end{eqnarray}
\begin{equation}
\widetilde{\mu}_{i} = 
\mu_{i} v_{5}^{2}  \int_{-\pi R}^{\pi R} dy \;
e^{-4\sigma} e^{2\sigma} \delta(y-\pi R)
= \mu_{i} v^{2}
\end{equation}
\begin{equation}
\widetilde{\zeta} = \zeta M_{pl} \int_{-\pi R}^{\pi R} dy 
e^{-4\sigma} e^{3\sigma} \delta(y-\pi R)
= \zeta M_{pl} e^{-\pi k R} = \zeta v
\end{equation}
\begin{eqnarray}
\widetilde{\chi}_{i} & = & \frac{\chi_{i}}{M_{pl}} \frac{1}{2\pi R} 
\left[
\frac{(2-2\alpha_{S})\pi kR}{e^{(2-2\alpha_{S})\pi kR} -1}
\right] 
\int_{-\pi R}^{\pi R} dy \; e^{-4\sigma} 
e^{2(2-\alpha_{S})\sigma} e^{2\sigma} 
\delta(y-\pi R) 
\nonumber\\
& = & \frac{\chi_{1}}{M_{pl}} \frac{1}{2\pi R} 
\left[
\frac{(2-2\alpha_{S})\pi kR}{e^{(2-2\alpha_{S})\pi kR} -1}
\right]
e^{(2-2\alpha_{S})\pi k R} \qquad (i = 1,2)
\end{eqnarray}
\begin{eqnarray}
\widetilde{\gamma}_{i} & = & \gamma_{i} \sqrt{M_{pl}} 
\frac{1}{\sqrt{2\pi R}}
\left[
\frac{(2-2\alpha_{S})\pi kR}{e^{(2-2\alpha_{S})\pi kR} -1}
\right]^{1/2}
\int_{-\pi R}^{\pi R} dy
e^{-4\sigma} e^{2\sigma} e^{(2-\alpha_{S})\sigma} \delta(y-\pi R)
\nonumber\\
& = & 
\gamma_{i} \sqrt{M_{pl}} 
\frac{1}{\sqrt{2\pi R}}
\left[
\frac{(2-2\alpha_{S})\pi kR}{e^{(2-2\alpha_{S})\pi kR} -1}
\right]^{1/2}
e^{-\alpha_{S} \pi k R} \qquad (i=1,2)
\end{eqnarray}
\begin{eqnarray}
\widetilde{\xi} & = &
\frac{\xi}{\sqrt{M_{pl}}} \frac{1}{\sqrt{2\pi R}}
\left[
\frac{(2-2\alpha_{S})\pi kR}{e^{(2-2\alpha_{S})\pi kR} -1} \right]^{1/2}
\; \int_{-\pi R}^{\pi R} dy 
e^{-4\sigma} e^{(2-\alpha_{S})\sigma} e^{3\sigma} \delta(y-\pi R)
\nonumber\\
& = & 
\frac{\xi}{\sqrt{M_{pl}}} \frac{1}{\sqrt{2\pi R}}
\left[
\frac{(2-2\alpha_{S})\pi kR}{e^{(2-2\alpha_{S})\pi kR} -1} \right]^{1/2}
e^{(1-\alpha_{S})\pi k R} \; .
\end{eqnarray}
As the effective coupling $\widetilde{\zeta} \sim v$ induces a weak scale 
triplet VEV, it is necessary to turn off $\zeta$. 
This can be done by imposing a $Z_{4}$ symmetry 
under which the fields transform as,
\begin{eqnarray}
H & \stackrel{Z_{4}}{\longrightarrow} & H
\\
\bigl(T, \, S \bigr) & \stackrel{Z_{4}}{\longrightarrow} & 
- \bigl(T, \, S \bigr)
\\
\Phi_{L} & \stackrel{Z_{4}}{\longrightarrow} & i \Phi_{L} \\
\Phi_{R} & \stackrel{Z_{4}}{\longrightarrow} & -i\Phi_{R}
\end{eqnarray}
This $Z_{4}$ symmetry forbids all the potentially 
dangerous trilinear couplings 
($\gamma_{1}, \, \gamma_{2},  \, \zeta$) in 
Eq.~(\ref{potential:vis}), and the 4D effective scalar 
potential is then given by,
\begin{eqnarray}\label{modeleff}
V_{eff} & = & 
\widetilde{\lambda}_{H} \widetilde{H}^{4}(x) 
+ \widetilde{\mu}_{H} \widetilde{H}^{2}(x) 
+\widetilde{\lambda}_{T} \widetilde{T}(x)^{4} 
+ \widetilde{\mu}_{T} \widetilde{T}(x)^{2} 
+ \widetilde{\lambda}_{S} S^{(0)}(x)^{4}
\nonumber\\
& & + \widetilde{\eta} \widetilde{T}(x)^{2} \widetilde{H}(x)^{2} 
+ \widetilde{\chi}_{1} \widetilde{T}(x)^{2} S^{(0)}(x)^{2}
+ \widetilde{\chi}_{2} \widetilde{H}(x)^{2} S^{(0)}(x)^{2}
\nonumber\\
&&
+ \widetilde{\xi} \widetilde{H}(x)\widetilde{T}(x)
\widetilde{H}(x)S^{(0)}(x) \; .
\end{eqnarray}
If the bulk singlet is localized close to the TeV brane, 
it then acquires a weak scale VEV, which leads to 
a large linear term for the triplet potential through the 
$\xi$ term, similar to the case with a bulk triplet discussed 
previously. Thus the bulk singlet must be localized close to 
the Planck brane, {\it i.e.} $\alpha_{S} > 1$. In this case, 
the effective couplings are given by,
\begin{eqnarray}
\widetilde{\chi}_{i} & \simeq & \chi_{i} 
\biggl(\frac{k}{M_{pl}}\biggr) (\alpha_{S}-1) 
\, e^{-2(\alpha_{S}-1)\pi k R}
\\
\widetilde{\xi} 
& \simeq & \xi \sqrt{\frac{k}{M_{pl}}} \sqrt{\alpha_{S}-1} 
\, e^{-(\alpha_{S}-1)\pi k R}
\; .
\end{eqnarray} 
By minimizing the effective potential given in 
Eq.(\ref{modeleff}), we obtain the following conditions,
\begin{eqnarray}
2\widetilde{\lambda}_{H} v^{2} + \bigl[
\widetilde{\mu}_{H}+\widetilde{\eta}u^{2}
+\widetilde{\chi}_{2}w^{2}+\widetilde{\xi}uw\bigr] &=&0
\\
2u\bigl[ 2\widetilde{\lambda}_{T} u^{2} 
+ \widetilde{\mu}_{T} + \widetilde{\eta} v^{2} 
+ \widetilde{\chi}_{1} w^{2} \bigr] + \widetilde{\xi}v^{2}w &=&0
\\
2w \bigl[ 2\widetilde{\lambda}_{S}w^{2} 
+ \widetilde{\chi}_{1}u^{2} + \widetilde{\chi}_{2}v^{2} 
\bigr] + \widetilde{\xi} v^{2} u &=&0
\end{eqnarray} 
where we have defined $<\widetilde{H}(x)> = v$, $<\xi^{0}(x)> = u$ and 
$<S^{(0)}(x)> = w$, with $\xi^{0}$ being the neutral component of the 
triplet, $\widetilde{T}(x)$. If $\chi_{1}>0$ and $\chi_{2} <0$,
assuming $v \gg w \gg u$ we obtain the following solutions, 
\begin{eqnarray}
v & = & \sqrt{\frac{-\widetilde{\mu}_{H}}{2\widetilde{\lambda}_{H}}}
= \sqrt{-\frac{\mu_{H}}{2\lambda_{H}}} v
\\
w & = & \sqrt{\frac{-\widetilde{\chi}_{2}v^{2}}{2\widetilde{\lambda}_{S}}}
= 
\sqrt{\frac{-\chi_{2}}{2\lambda_{S}} \frac{k}{M_{pl}}}\sqrt{(\alpha_{S}-1)}
v e^{-(\alpha_{S}-1) \pi k R}
\label{wvev}
\\
u & = & -\frac{\widetilde{\xi} v^{2}w}
{2\widetilde{\mu}_{T}+2\widetilde{\eta}v^{2}} 
= -\frac{1}{2(\mu_{T}+\eta)} \xi \sqrt{\frac{-\chi_{2}}{2\lambda_{S}}} 
\bigl(\alpha_{S}-1\bigr) 
\biggl(\frac{k}{M_{pl}}\biggr) v e^{-2(\alpha_{S}-1) \pi kR} \; .  
\label{uvev}
\end{eqnarray}
As we can see the assumption $v \gg w \gg u$ is justified. 
A crucial point to note is that, 
in order for $w$ given in Eq.~(\ref{wvev}) to be the true vacuum, 
$\chi_{2}$ must be 
negative. Otherwise the VEV will be determined by the linear 
term of $S$, leading a large VEV 
for the singlet. Similarly, in order for $u$ given in Eq.~(\ref{uvev}) 
to be the true vacuum, $\chi_{1}$ must be positive. 
Otherwise the VEV will be determined by the quartic and quadratic 
terms of $T$ 
in the potential, leading to a large VEV for the triplet.

The Yukawa coupling for the charged fermions to the Higgs doublet reads
\begin{equation}\label{yuk}
\frac{Y_{ij}}{M_{pl}} \int d^{4}x \int dy \sqrt{-g}
\overline{\Psi}_{R,i}(x,y) \Psi_{L,j}(x,y) H(x) \delta(y-\pi R)  \; ,
\end{equation}
where $Y_{ij}$ are dimensionless $\mathcal{O}(1)$ coefficients. 
The $SU(2)$ Higgs doublet is assumed to be confined to the 
TeV brane as we have 
explained in Sec.~\ref{introduction}. 
The effective Yukawa coupling in four dimensions is obtained after
integrating out the fifth coordinate, $y$:
\begin{eqnarray}
\widetilde{Y}_{ij} & = & \frac{Y_{ij}}{M_{pl}}
(\frac{1}{2\pi R})
\biggl[
\frac{(1-2c_{R,i})\pi kR}{e^{(1-2c_{R,i})\pi kR} -1 }
\biggr]^{1/2}
\biggl[
\frac{(1-2c_{L,j})\pi kR}{e^{(1-2c_{L,j})\pi kR} -1 }
\biggr]^{1/2}
\nonumber\\
& & \qquad \quad \cdot 
\int_{-\pi R}^{\pi R} dy \sqrt{-g} \;
e^{(2-c_{R,i})\sigma} e^{(2-c_{L,j})\sigma}
e^{\sigma} \delta(y-\pi R)
\nonumber\\
& = & \frac{Y_{ij}}{2} \frac{k}{M_{pl}} 
\frac{\sqrt{(1-2c_{R,i})(1-2c_{L,j})}}
{\sqrt{(e^{(1-2c_{R,i})\pi kR}-1)(e^{(1-2c_{L,i})\pi kR}-1)}}
e^{(1-c_{R,i} - c_{L,j})\pi k R} \; .
\end{eqnarray}
The interaction between the leptons and the triplet gives the Majorana 
neutrino masses,  
\begin{equation}
\frac{\lambda_{ij}}{M_{pl}} \int d^{4}x \int dy \sqrt{-g}
\nu^{T}_{L,i}(x,y) \nu_{L,j}(x,y) \xi^{0}(x) \delta(y-\pi R) \; .
\end{equation}
Because the triplet is also confined to the TeV brane, the effective 
coupling of the leptons to the triplet is of the same form as 
that to the SM Higgs doublet, 
\begin{equation}
\widetilde{\lambda}_{ij} = \frac{\lambda_{ij}}{2} \frac{k}{M_{pl}} 
\frac{\sqrt{(1-2c_{L,i})(1-2c_{L,j})}}
{\sqrt{(e^{(1-2c_{L,i})\pi kR}-1)(e^{(1-2c_{L,i})\pi kR}-1)}}
e^{(1-c_{L,i} - c_{L,j})\pi k R} \; .
\end{equation}
In a large parameter space, $\widetilde{\lambda}_{ij}$ and 
$\widetilde{Y}_{ij}$ are of the 
same order. Thus small neutrino Majorana masses (see Eq.~(\ref{triplet})) 
are obtained because $\bigl< \xi^{0}_{(0)}(x) \bigr>$ is highly suppressed 
relative to $\bigl<\widetilde{H}(x) \bigr>$ due to the warped geometry.

\section{Numerical results}\label{result}

In order to analyze the phenomenological implications of 
our model, we define 
\begin{equation}
\epsilon(x) \equiv \frac{\sqrt{x}}{\sqrt{e^{x\pi kR}-1}}, \qquad
\eta(x) \equiv e^{-x \pi k R} \; ,
\end{equation}
and write the coupling constants, $\widetilde{\lambda}_{ij}$ 
and $\widetilde{Y}_{ij}$, as
\begin{eqnarray}
\widetilde{\lambda}_{ij} & = &
\frac{\lambda_{ij}}{2} \frac{k}{M_{pl}} 
\epsilon(1-2c_{L,i})\epsilon(1-2c_{L,j})\eta(c_{L,i})
\eta(c_{L,j}) e^{\pi kR}
\\
\widetilde{Y}_{ij} & = &
\frac{Y_{ij}}{2} \frac{k}{M_{pl}} 
\epsilon(1-2c_{R,i})\epsilon(1-2c_{L,j})\eta(c_{R,i})
\eta(c_{L,j}) e^{\pi kR} \; .
\end{eqnarray}
Without imposing any additional symmetry, the $5D$ Yukawa couplings 
to the triplet for 
all the three families are universal, that is, $\lambda_{ij}=1$ 
for all $(ij)$.  
In this case, the neutrino mass matrix arising from this mechanism 
has the nearest neighbor structure, 
and to the leading order has the following form, 
\begin{equation}
M_{\nu} \sim
\left(\begin{array}{ccc}
t^{2} & t & t 
\\
t & 1 & 1
\\
t & 1 & 1
\end{array}
\right) \; ,
\end{equation}
where we have set $c_{L,\mu} = c_{L,\tau}$, 
which is implied by the maximal mixing angle in the atmospheric neutrino 
sector, 
$\sin^{2}2\theta_{\mu\tau}=1.00$~\cite{Fukuda:2000np}. A neutrino mass 
matrix of this form, nontheless,  always gives a solar mixing angle 
$\tan^{2}\theta_{\odot} > \pi/4$, while the presence of the matter 
effect requires the solar mixing angle to be in the dark side, 
$\tan^{2}\theta_{\odot} < \pi/4$~\cite{deGouvea:2000cq}. 
We thus impose an $Z_{4}$ symmetry, under which, the fermions 
transform in the following way,
\begin{eqnarray}
\left( \Phi_{L,\mu}, \; \Phi_{R,e}, \; \Phi_{R,\tau} \right)
& \stackrel{Z_{4}^{\prime}}{\longrightarrow} & \; (i) 
\left( \Phi_{L,\mu}, \; \Phi_{R,e}, \; \Phi_{R,\tau} 
\right)
\label{z41}\\
\left( \Phi_{L,e}, \; \Phi_{L,\tau}, \; \Phi_{R,\mu} \right)
& \stackrel{Z_{4}^{\prime}}{\longrightarrow} & \; (-i)
\left( \Phi_{L,e}, \; \Phi_{L,\tau}, \; \Phi_{R,\mu} \right)
\; .
\label{z42}
\end{eqnarray}
The scalar fields are neutral under this $Z_{4}^{\prime}$ symmetry. 
This $Z_{4}^{\prime}$ symmetry is softly broken in the bulk 
through the Yukawa couplings to the triplet scalar field,
\begin{equation}\label{lambda5}
\lambda_{ij}=\left(\begin{array}{ccc}
x & 1 & x\\
1 & x & 1  \\
x & 1 & x
\end{array}\right) \; .
\end{equation}
where the parameter $x < 1$ characterizes the amount 
of breaking, and the neutrino mass matrix is then 
given by $(M_{\nu})_{ij} = 
\widetilde{\lambda}_{ij} m_{\nu}$,
where the overall mass scale $m_{\nu}$ is,
\begin{equation}
m_{\nu} = \frac{1}{2(\mu_{T}+\eta)} \xi \sqrt{\frac{-\chi_{2}}{2\lambda_{S}}} 
\bigl(\alpha_{S}-1\bigr) 
\biggl(\frac{k}{M_{pl}}\biggr) v e^{-2(\alpha_{S}-1) \pi kR}
 \; .
\end{equation}
The bounds from the electroweak precision measurements on the fermionic 
mass parameters 
are rather weak for $c \ge 1/2$; however, the 
constraints grow strong rapidly 
for $c < 1/2$~\cite{delAguila:2000kb}. 
As a result, we consider only the region $c \ge 1/2$. 
Clearly, many possible sets of  
bulk mass parameters can be chosen to accommodate the 
observed fermion masses and 
mixing angles. Let us assume that both the $\mu$ and 
$\tau$ lepton doublets are 
de-localized, that is, $c_{L,\mu} = c_{L,\tau} = 1/2$. 

We consider $k \simeq M_{pl}$ and $kR = 11$ which 
translates into a warp factor 
$e^{-\pi kR} = 9.8 \times 10^{-16}$. 
The dimensionless parameters in the scalar potential 
are chosen to be $\bigl( \lambda_{S},\mu_{T},\eta,
\xi,\chi_{2}\bigr)
=\bigl( 1/2,1/2,1/2,-1,-1 \bigr)$. 
The rest of the parameters in the scalar sector 
are irrelevant, as long as they are of $\mathcal{O}(1)$.
With the choice $c_{L,\mu} = c_{L,\tau} = 1/2$, 
using $\Delta m_{atm}^{2} \equiv m_{3}^{2} - m_{2}^{2} 
\simeq m_{3}^{2} = 2.3 \times 10^{-3} \; eV^{2}$ as an input, 
the bulk scalar mass parameter is determined to be $\alpha_{S}=1.35$,  
which corresponds to a bulk mass 
term $a_{S} k^{2} = -2.18 k^{2}$. The VEV of the neutral component of the 
triplet is highly suppressed,  
$\bigl<\xi^{(0)}(x)\bigr> \simeq  
(1-\alpha_{S})
e^{-2(\alpha_{S}-1)\pi kR} v \simeq 1.11 \times 10^{-11} v$,  
where the VEV of the SM doublet, $v$, is $v=174$ GeV. 
At the first glance one might think that this bulk 
mass term is larger than the $5D$ Planck scale. This turns 
out not to be the case. 
In order to derive the metric given in Eq.~(\ref{RSmetric}), 
one has to assume that the $5$-dimensional curvature scalar, 
$R_{5} = -20k^{2}$, satisfies $|R_{5}| < M_{pl}^{2}$~\cite{Randall:1999ee}. 
This condition translates into an upper bound on $k$. 
It can be seen easily that, by slightly lowering the value of $k$, the 
condition $\bigl| a_{S}k^{2} \bigr| < \bigl| R_{5} \bigr| < M_{pl}^{2}$ 
is satisfied, and thus our result can be trusted.  

\begin{figure}[t]
\psfrag{e_L}[][]{$e_{L}$}
\psfrag{mu_L}[][]{$\mu_{L}$, $\tau_{L}$}
\psfrag{e_R}[][]{$e_{R}$}
\psfrag{mu_R}[][]{$\mu_{R}$}
\psfrag{tau_R}[][]{$\tau_{R}$}
\psfrag{s}[][]{$S$}
\psfrag{pi*y}[][]{$k y / \pi$}
\psfrag{f_0}[][]{$e^{-z\sigma} f_{0}$}
\includegraphics[scale=0.4,angle=270]{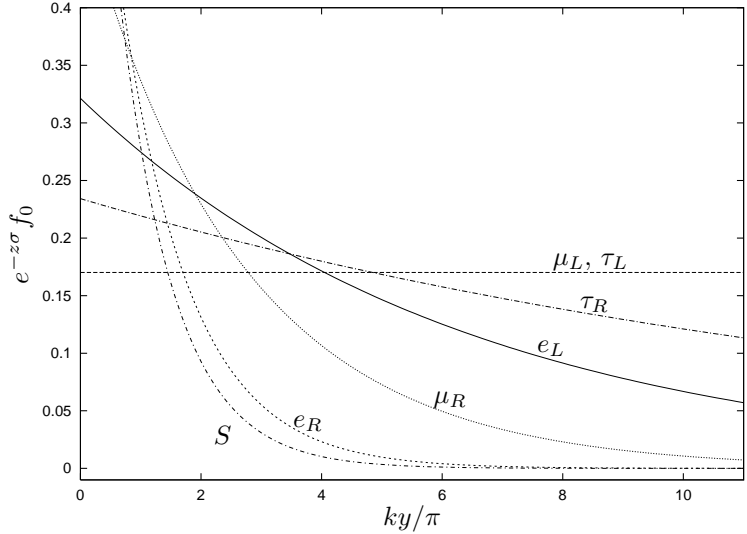}
\caption{\label{cm} The profile for the zero mode of various bulk fields 
with corresponding values of $c$-parameters considered in the model. 
Taking into account the measure due to the warp geometry, we have 
$z=1/2$ for fermions, and $z=-1$ for bosons. In this case the triplet 
and the SM Higgs doublet are both confined to the TeV brane.}
\end{figure}

Using the LMA solution, $\Delta m_{\odot}^{2} 
\equiv m_{2}^{2} - m_{1}^{2} 
= 8.0 \times 10^{-5} \; eV^{2}$ as an 
input~\cite{Ahmad:2002ka,Ahmad:2002jz,Bahcall:2004ut,Maltoni:2004ei}, 
we find,
\begin{equation}
c_{L,e} = 0.55 \; .
\end{equation} 
With $x=0.797$, the neutrino mass matrix is then given by
\begin{equation}
M_{\nu} = \left(
\begin{array}{ccc}
0.00130 & 0.00486 & 0.00387 \\
0.00486 & 0.0115 & -0.0145 \\
0.00387 & -0.0145 & 0.0115
\end{array}\right)\cdot m_{\nu} \; , \qquad 
\mbox{where} \; \; m_{\nu} = 
\biggl< \xi^{0} (x) \biggr> = 1.11 \times 10^{-11} v \;.
\end{equation}
These parameters give
\begin{equation}
m_{1} = 0.0110 \; eV, \quad m_{2} = 0.0142 \; eV, 
\quad m_{3} = 0.0501 \; eV \; , 
\end{equation}
and the full neutrino mixing matrix reads
\begin{equation}
V_{\alpha i} \simeq  
\left(\begin{array}{ccc}
0.813 & -0.581 & 0.0298\\
0.392 & 0.584 & 0.711\\
0.431 & 0.567 & -0.702
\end{array}\right) \; ,
\end{equation}
where $\alpha=e,\mu,\tau$ (flavor eigenstates) 
and $i=1,2,3$ (mass eigenstates).
The predictions with these parameters are
\begin{eqnarray}
\tan^{2} \theta_{\odot} & = & 0.512 \; 
(\mbox{agrees with LMA solution~\cite{Bahcall:2004ut,Maltoni:2004ei}}),
\qquad \label{p1}\\
|U_{e\nu_{3}}| & \sim & 0.0298 \; 
(\mbox{agrees with CHOOZ 
$1 \sigma$ bound $|U_{e\nu_{3}}| < 
0.12$~\cite{Apollonio:1999ae,Bahcall:2004ut,Maltoni:2004ei}}) \; .
\label{p2}
\end{eqnarray}
Now we consider the charged lepton sector. 
The Yukawa coupling is dictated by the $Z_{4}$ charge 
assignment given in Eq.\ref{z41} and \ref{z42}; it is, 
\begin{equation}
Y_{e} = \left(
\begin{array}{ccc}
1 & 0 & 1\\
0 & 1 & 0\\
1 & 0 & 1
\end{array}\right)
\end{equation}
Using the charged lepton masses~\cite{Hagiwara:2002pw}
\begin{equation}
m_{e} = 0.504 \; MeV, 
\quad
m_{\mu} = 108 \; MeV,
\quad
m_{\tau} = 1.77 \; GeV \; , 
\end{equation}
we found the following solutions for $c_{R}$:
\begin{equation}
c_{R,e} = 0.776, \quad
c_{R,\mu} = 0.622, \quad
c_{R,\tau} = 0.521 \; ,
\end{equation}
and the corresponding charged lepton mass matrix is given by,
\begin{equation}
M_{e} = \left(\begin{array}{ccc}
-2.66 \times 10^{-4} & 0 & 7.92 \times 10^{-4}\\
0 & 0.108 & 0\\
0.563 & 0 & 1.68\end{array}\right) \; \mbox{GeV}
\; . 
\end{equation}
The $1-3$ mixing in the charged lepton mass matrix is 
of the order of $\mathcal{O}(10^{-4})$, which is negligibly small and 
do not affect the predictions given in Eq.~\ref{p1} and \ref{p2} 
to the accuracy we are considering. 
This set of $c$-parameters give rise to the profiles of the zero mode of 
various bulk fields as shown in Fig.~\ref{cm}.

We comment that the mass of the $SU(2)$-triplet 
scalar field is of the order of a TeV, 
which is substantially lower than that in the usual 
triplet mechanism implemented in $4D$~\cite{Ma:1998dx}. 
In the usual $4D$ case, the triplet mass is required to be 
of the order of $10^{13} GeV$. The difference between 
these two cases arises because the tri-linear 
coupling constant between the SM Higgs doublet and the $SU(2)$ triplet 
in the usual $4D$ case is of the order of the weak scale,  
while it is suppressed by the warped factor in  
our case. The triplet with a TeV scale mass 
could be produced at the upgraded Tevatron 
or at the Large Hadron Collider. By measuring 
the decay branching ratio of 
\begin{equation}
\xi^{++} \rightarrow l_{i}^{+} l_{j}^{+}
\end{equation}
it is possible to map out the whole triplet 
coupling matrix, $\widetilde{\lambda}_{ij}$, and thus 
the mixing matrix in the leptonic sector, making this model verifiable.
The presence of this $SU(2)$-triplet Higgs might also have  
interesting implication for leptogenesis 
through the decay of the doubly charged component of the 
triplet scalar field~\cite{Chen:2004ww}.
The TeV-brane induced mass of the singlet zero mode, 
on the other hand, is suppressed by 
$e^{-(\alpha_{S}-1)\pi kR}$ relative 
to the weak scale, due to the small overlap between 
its wave function and the SM Higgs. 
The induced mass of the singlet zero mode is closely linked to the 
absolute mass scale of the neutrinos.  
For $\alpha_{S}=1.35$, it is  
$\sim v e^{-(\alpha_{S}-1)\pi kR} \simeq 0.5$ MeV. As 
the singlet does not have any SM interactions (it only couples 
weakly to the SM Higgs and the triplet through 
the couplings $\chi_{i}$ and $\xi$), there are no experimental 
constraints on the mass of the singlet. 
%And because it is weakly 
%interacting, this singlet is a good candidate for 
%the dark matter. We leave this possibility for 
%further investigation~\cite{cm5}.

\section{Conclusions}\label{conclude}

In this paper we propose a new mechanism which naturally 
gives rise to small neutrino Majorana masses in $5D$ with 
warped geometry. This is realized at tree level by coupling 
the lepton doublets to the $SU(2)$ triplet scalar field.  
The smallness of neutrino masses is due to 
the small overlap between the profile of a bulk singlet zero mode, 
which is localized close to the Planck brane, and the TeV-brane confined 
SM Higgs and the triplet. 
We emphasize that even though this mechanism 
is not predictive in the sense that it does not 
reduce the number of parameters 
in the Yukawa sector, it provides a way to generate large mass hierarchy 
from parameters that are all naturally of $\mathcal{O}(1)$. 
The generic form for the neutrino mass matrix due to the 
overlap between the fermions is not compatible with the 
LMA solution. This is overcome by imposing a $Z_{4}$ symmetry, 
which is softly broken by the couplings of the triplet Higgs
 to the lepton doublets.  
This model successfully reproduces the observed masses
 and mixing angles in charged lepton sector as well 
as in the neutrino 
sector, in addition to having a prediction of 
small $\left| U_{e3} \right| \sim \mathcal{O}(0.01)$, which 
is in the range accessible to the future long baseline neutrino 
experiments~\cite{Diwan:2003bp}. 
As the mass of the triplet scalar field is of 
the TeV scale, it could be produced at the upcoming 
collider experiments. Once it is produced, 
by measuring the decay branching ratio of 
$\xi^{++} \rightarrow l_{i}^{+} l_{j}^{+}$, 
it is possible to map out the whole triplet coupling matrix, 
thus the mixing matrix in the neutrino sector.
%The mass of the singlet zero mode is of the order of a MeV. 
%As the singlet is weakly interacting, it is a natural candidate for 
%the dark matter.

\begin{acknowledgments}
This work was supported, in part, by the U.S. Department 
of Energy under grant number DE-AC02-98CH10886. I  
thank Ryu Kitano and K.T. Mahanthappa for 
helpful conversations.
\end{acknowledgments}

\bibliography{ed3}

%}

\end{document}